\documentclass{article}%
\usepackage{amsmath}
\usepackage{amsfonts}
\usepackage{amssymb}
\usepackage{graphicx}%
\providecommand{\U}[1]{\protect\rule{.1in}{.1in}}

\begin{document}
\begin{titlepage}
\vspace{.3cm} \vspace{1cm}
\begin{center}
\baselineskip=16pt \centerline{\Large\bf  Quantum Cosmological Perturbations: Predictions and Observations } \vspace{2truecm} \centerline{\large\bf
\ Viatcheslav Mukhanov\ \ } \vspace{.5truecm}
\emph{\centerline{Theoretical Physics, Ludwig Maxmillians University,Theresienstr. 37, 80333 Munich, Germany }}
\emph{\centerline{LPT de l'Ecole Normale Superieure, Chaire Blaise Pascal, 24 rue Lhomond, 75231 Paris cedex, France}}
\end{center}
\vspace{2cm}
\begin{center}
{\bf Abstract}
\end{center}
I consider the generic model independent predictions of the theory of quantum cosmological perturbations. To describe the stage of cosmic inflation, where these perturbations are amplified,
the hydrodynamical approch is used. The inflationary stage is completely characterized by the deviation of the equation of state from cosmological constant
which is a smooth function of the number of e-folds until the end of inflation. It is shown that in this case the spectral index should deviate from the flat one at least
by 3 percent irrespective of any particular scenario. Given the value of the spectral index the lower bound on the amount of the gravitational
waves produced is derived. Finally the relation between effective hydrodynamical description of inflation and inflationary scenarios is discussed.
\end{titlepage}

\section{\bigskip Introduction}

It was discovered in 1980 that the quantum fluctuations of the metric can
explain the observable structure of the universe \textit{if and only if the
expanding universe} went through a stage of cosmic acceleration \cite{MC1}.
The spectrum of these perturbations in the range of observable scales was
calculated for the first time in \cite{MC2}. About the same time it was
realized that in order to understand the large scale homogeneity and isotropy
of the observable universe one also needs a stage of accelerated expansion,
cosmic inflation \cite{GL}. At present there exist hundreds of different
inflationary scenarios. To understand what theory of quantum fluctuations
really predicts and how to extract the parameters, characterizing the
inflationary stage, from observations it is convenient to describe inflation
using the effective hydrodynamical approach. In this approach the state of the
matter is entirely characterized by its energy density $\varepsilon$ and the
pressure~$p$. In this note I will only consider the predictive inflationary
theory, when both the acceleration and the perturbations are due to the same
matter component. There exist many models where one kind of matter is
responsible for acceleration and the other for perturbations. In these models
nearly any outcome of the measurements can be accommodated making them
experimentally non falsifiable and therefore of no great interest. Indeed a
theory makes sense only if it makes non-trivial predictions which can be
confirmed or disproved by measurements and the best theory is the theory with
the minimal number of parameters. In fact, there is no need to involve more
parameters unless there appears an obvious contradiction with experimental
data or there exist deep theoretical reasons for doing so. Because the theory
of simple inflation is in excellent agreement with the present observations it
is enough to restrict ourselves to this predictive theory.

\section{Robust predictions}

We assume that in the past the universe went through a stage when matter with
equation of state $p\approx-\varepsilon$ was dominating and, hence, the
universe was accelerating. A cosmological constant corresponding to
$p=-\varepsilon$ cannot serve our purpose because finally one has to have a
graceful exit from inflation. Therefore, from the very beginning there should
be small deviations of the equation of state from the cosmological constant,
i.e., $\left(  \varepsilon+p\right)  /\varepsilon\ll1,$ \textit{but
nonvanishing}. This ratio, which at the beginning should be small enough to
provide us the necessary duration of inflation, grows until it becomes of
order unity when inflation ends through a graceful exit to decelerated
expansion. One can realize the needed equation of state using the condensates
of scalar fields, the $R^{2}$ gravity and in some other ways. The key point is
that the microscopic origin of the dark energy does not play a crucial role
regarding the major predictions of the quantum cosmological perturbations
theory. Everything we need is a \textquotedblleft decaying cosmological
constant\textquotedblright.

To describe how this \textquotedblleft cosmological constant\textquotedblright%
\ decays we will use as a time parameter the number of e-fold $N$ left to the
end of inflation, defined as
\begin{equation}
a=a_{f}\exp\left(  -N\right)  , \label{1}%
\end{equation}
where $a$ is the scale factor and $a_{f}$ is its value at the end of inflation
when $\left(  \varepsilon+p\right)  /\varepsilon\simeq O\left(  1\right)  .$
For observations the relevant interval of $N$ is not very large, namely,
$N<70.$ Moreover, the spectrum of fluctuations observed in CMB corresponds to
even smaller interval, $N\simeq50-60.$ Making the reasonable assumption that
during this rather short range of $\ N$ the change of the equation of state is
monotonic and smooth and taking into account that $\left(  \varepsilon
+p\right)  /\varepsilon\simeq O\left(  1\right)  $ at $N=0,$ it is rather
natural to approximate the equation of state by
\begin{equation}
1+\frac{p}{\varepsilon}=\frac{\beta}{\left(  N+1\right)  ^{\alpha}}, \label{2}%
\end{equation}
where $\alpha$ and $\beta$ are both positive and of order unity. \ Within this
set up, irrespective of the initial conditions for the
perturbations\footnote{It is often mistakenly stated in the literature that
one has to postulate the initial "Bunch-Davies vacuum" for the cosmological
perturbations. In fact assuming that duration of inflation lasts longer than
70 e-folds the spectrum generated in the observable scales does not depend on
the initial conditions for the perturbations provided that they do not destroy
via backreaction the inflationary stage from the very beginning.}, concrete
robust predictions for observations can be derived. What are these predictions?

First of all if inflation last more than 70 e-folds the cosmological parameter
$\Omega$ should be equal to unity\textit{\ }within an accuracy about
$10^{-5},$ which means that at present the universe has a flat Euclidean
geometry. This prediction was first confirmed only at the end of 90th with the
discovery of dark energy. Note that the experimental data before were in
strong disagreement with it.

The other set of the robust predictions concerns the amplified quantum
fluctuations \cite{MC2}. More concretely:

-The produced inhomogeneities should be adiabatic. I would like to stress that
about thirty years ago the observational data were more supportive for entropy
perturbations. However, nowadays they are ruled out by the precision CMB
measurements, which confirmed the adiabatic nature of the primordial inhomogeneities.

-The primordial inhomogeneities are nearly Gaussian. This is because they were
originated as the result of amplification by the external classical
gravitational field of the initial gaussian fluctuations. The expected
corrections to the gaussian gravitational potential $\Phi_{g},$ due to
nonlinear corrections to the linearized Einstein equations are of order
$O\left(  1\right)  \Phi_{g}^{2}$, that is, $\Phi=\Phi_{g}+f_{NL}\Phi_{g}%
^{2}.$ The present experimental bound $-10<f_{NL}<70$ is in agreement with the
prediction of the theory, according to which $f_{NL}$ is expected to be about
ten. The forecasted accuracy of the Planck mission $\Delta f_{NL}\simeq5$ will
allow us to improve further the measurements of the non-gaussianity.

-The most nontrivial prediction for the perturbations is a weak scale
dependence of the amplitude of the gravitational potential $\Phi$. Namely, the
amplitude of $\Phi$ must logarithmically depend on the scale $\lambda$ and
grow towards the larger scales. Within the observable range of scales the
logarithm can be approximated by $\Phi^{2}\propto\lambda^{1-n_{s}}.$ The
rather generic formula for the spectral index $n_{s}$ is \cite{Mbook}:%
\begin{equation}
n_{s}-1=-3\left(  1+\frac{p}{\varepsilon}\right)  +\frac{d}{dN}\ln\left(
1+\frac{p}{\varepsilon}\right)  . \label{3}%
\end{equation}
For a monotonic change of the equation of state both terms are negative and
hence inflation \textit{always} predicts the red-tilted spectrum. Substituting
(\ref{2}) into (\ref{3}) we find that
\begin{equation}
n_{s}-1=-\frac{3\beta}{\left(  N+1\right)  ^{\alpha}}-\frac{\alpha}{\left(
N+1\right)  } \label{4}%
\end{equation}
For $\alpha<1$ the first term on the r.h.s. dominates, while for $\alpha>1$
the second term is more important. In case of $\alpha=1$ both terms give a
comparable contribution. Considering $n_{s}-1$ as a function of $\alpha$ we
find that the minimal deviation from the flat spectrum ($n_{s}=1$) corresponds
to
\begin{equation}
\alpha=1+\frac{\ln\left(  3\beta\ln\left(  N+1\right)  \right)  }{\ln\left(
N+1\right)  }. \label{5}%
\end{equation}
Substituting this expression in (\ref{4}) we will find that inflation predicts
that the deviation of the spectral index from unity must be larger than
\begin{equation}
n_{s}-1=-\left(  1+\frac{\ln\left(  3\beta e\ln\left(  N+1\right)  \right)
}{\ln\left(  N+1\right)  }\right)  \frac{1}{\left(  N+1\right)  }. \label{6}%
\end{equation}
Taking $3\beta=1$ ($\beta$ cannot be much smaller that unity) and noting that
$N=50$ for the scales where this spectral index is measured, we find that the
predicted spectral index should be smaller than $n_{s}=0.968.$ The logarithmic
spectrum obtained in \cite{MC2} corresponds to $n_{s}=0.96$ and this
prediction is in good agreement with the most recent measurements of the CMB
fluctuations \cite{Bond}, which give $n_{s}=0.9690\pm0.0089$ and confirm the
logarithmic dependence of the gravitational potential at the level of
3,5$\sigma.$ This logarithmic dependence has a deep physical origin since it
is due to the small deviation of the equation of state from cosmological
constant needed for a graceful exit.

If any of the above predictions would contradict to the observations then
inflation as a predictive theory (sometimes called simple inflation) would be
ruled out.

-One more robust prediction of inflation is the existence of the longwave
gravitational waves \cite{Star1}. The ratio of the tensor ($T$) to scalar mode
($S$) can also be expressed in terms of the equation of state (see
\cite{Mbook}):
\begin{equation}
r=\frac{T}{S}=24\left(  1+\frac{p}{\varepsilon}\right)  =\frac{24\beta
}{\left(  N+1\right)  ^{\alpha}} \label{7}%
\end{equation}
One can immediately see that taking, for instance, $\alpha=2$ we can reduce
the amount of the longwave gravitational waves by a factor $N=50$ compared to
the case $\alpha=1.$ Hence, the particular value of the ratio $r$ is not
predicted by generic theory of inflation. However, after spectral index
$n_{s}$ is measured one can establish most likely lower bound on the amount of
the generated gravitational waves. This bound does not depend on the
particular inflationary scenario. In fact, for $\alpha>1,$
\begin{equation}
n_{s}-1\simeq-\frac{\alpha}{N}, \label{7a}%
\end{equation}
and if, for instance, the measured value $n_{s}\simeq0.96,$ then $\alpha$
cannot be larger than $2$ and, hence, the lower bound on $r$ is%
\begin{equation}
r=\frac{24\beta}{\left(  N+1\right)  ^{2}}\simeq10^{-2}\beta, \label{7c}%
\end{equation}
where $\beta$ is of order unity. For the spectral index $n_{s}\simeq0.94$ the
amount of the gravity waves can be further suppressed by a factor 50. Hence,
the non-detection of the gravitational waves in the current CMB measurements
does not rule out the predictive inflationary theory, but on the other hand
their detection would \ provide an extra strong evidence for inflation.

I would like to stress that the model independent predictions above are
extremely nontrivial and were for a long time in conflict\ with observations.
For example, in the 80th, along with the theory of quantum initial
perturbations there were competing theories of cosmic strings, textures and
entropy perturbations, which sometimes were even more favorable from the point
of view of observations. However, now all these theories are ruled out and
only the theory of quantum cosmological perturbations with all its nontrivial
predictions is confirmed by observations. Moreover, although there are still
claims in the literature that there are alternatives to inflation, there is no
any alternative to the quantum origin of the universe structure.

After the origin of the universe structure from quantum fluctuations is
confirmed one can ask the question how much can we really learn about
fundamental physics making precise measurements of the parameters $\alpha$ and
$\beta.$

\section{\bigskip Slow roll inflation}

Assuming that inflation is due to the slow roll scalar field with a standard
kinetic energy term we will determine the scalar field potentials, which
correspond to different values of $\alpha$ and $\beta$. Keeping in mind that
the energy density during inflation is approximately equal to the potential,
that is, $\varepsilon\simeq V\left(  \varphi\right)  \,,$ we first determine
how the energy density depends on $N.$ The energy conservation equation%
\begin{equation}
\dot{\varepsilon}=-3H\left(  \varepsilon+p\right)  , \label{8}%
\end{equation}
where $H=\dot{a}/a$ and the dot denotes the time derivative, can be rewritten
as%
\begin{equation}
\frac{d\ln\varepsilon}{dN}=3\left(  1+\frac{p}{\varepsilon}\right)
=\frac{3\beta}{\left(  N+1\right)  ^{\alpha}}. \label{9}%
\end{equation}
Integrating this equation we obtain%
\begin{equation}
\varepsilon\left(  N\right)  \simeq\left\{
\begin{array}
[c]{c}%
\varepsilon_{f}\left(  N+1\right)  ^{3\beta},\text{
\ \ \ \ \ \ \ \ \ \ \ \ \ \ \ \ \ \ \ \ \ }\alpha=1,\\
\varepsilon_{0}\exp\left(  -\frac{3\beta}{\alpha-1}\frac{1}{\left(
N+1\right)  ^{\alpha-1}}\right)  ,\text{ \ \ \ \ \ }\alpha\neq1.
\end{array}
\right.  \label{10}%
\end{equation}
To determine the slow roll potential $V\left(  \varphi\right)  \simeq
\varepsilon$ we have to express $N$ in terms of the scalar field $\varphi.$
With this purpose we write%
\begin{equation}
\frac{d\varphi}{dN}=\frac{\dot{\varphi}}{-H}=\sqrt{\frac{3}{8\pi}\left(
1+\frac{p}{\varepsilon}\right)  }, \label{11}%
\end{equation}
where we have taken into account that $\varepsilon+p=\dot{\varphi}^{2}$ and
$H^{2}=8\pi\varepsilon/3$ (in the Planck units). Substituting here (\ref{2})
and integrating the resulting equation we find
\begin{equation}
N+1=\left\{
\begin{array}
[c]{c}%
\exp\left(  \pm\sqrt{\frac{8\pi}{3\beta}}\left(  \varphi+C\right)  \right)
,\text{ \ \ \ \ \ \ \ \ \ \ }\alpha=2,\\
\left[  \frac{2\pi}{3\beta}\left(  2-\alpha\right)  \right]  ^{\frac
{1}{2-\alpha}}\left(  \pm\varphi+C\right)  ^{\frac{2}{2-\alpha}},\text{
}\alpha\neq2,
\end{array}
\right.  \label{12}%
\end{equation}
where $C$ is a constant of integration.

Let us consider the cases $\alpha=1,$ $\alpha=2$ and $\alpha\neq1,2$ separately.

- $\alpha=1$

Taking into account that during slow roll $V\left(  \varphi\right)
\simeq\varepsilon$ and combining (\ref{10}) and (\ref{12}) we find that
\begin{equation}
V\left(  \varphi\right)  \simeq V\left(  \varphi_{f}\right)  \left(
\frac{\varphi}{\varphi_{f}}\right)  ^{6\beta}, \label{13}%
\end{equation}
where $\varphi_{f}=\left(  3\beta/2\pi\right)  ^{1/2}.$ Thus, the case of
$\alpha=1$ corresponds to chaotic inflationary scenarios with the power law
potentials\cite{Linde2}. In this case both terms in (\ref{3}) give comparable
contributions to spectral index, which becomes%
\begin{equation}
n_{s}-1=-\frac{3\beta+1}{N+1}, \label{14}%
\end{equation}
and the amount of the gravitational waves produced is rather substantial%
\begin{equation}
r=\frac{24\beta}{N+1}. \label{15}%
\end{equation}
For instance, for the massive scalar field ($3\beta=1)$ we have $n_{s}%
\simeq0.96$ and $r\simeq0.16$ for $N=50,$ while in the case $3\beta=2,$ that
is, for quartic potential $n_{s}\simeq0.94$ and $r\simeq0.32.$ The measured
value of the spectral index $n_{s}=0.9690\pm0.0089$ and the upper bound on the
amount of the gravitational waves seems disfavor the quartic potential at
rather high significance level. The model of massive scalar field predicts the
spectral index to be in agreement with observations, but the amount of the
produced gravitational waves seems too high although one cannot yet definitely
rule out this model \cite{komatsu}.

-$\alpha=2$

In this case the second term in (\ref{3}) dominates. The spectral index does
not significantly depends on $\beta$ and is equal to%
\begin{equation}
n_{s}-1\simeq-\frac{2}{N}. \label{16}%
\end{equation}
For $N=50$ the spectral index $n_{s}\simeq0.96$ and the amount of the
gravitational waves $r\simeq10^{-2}\beta$ are both in very agreement with
observations. As it follows from (\ref{10}) and (\ref{12}) this case
corresponds to the following slow roll potential%
\begin{equation}
V\left(  \varphi\right)  \simeq V_{0}\exp\left(  -3\beta\exp\left(  \mp
\sqrt{\frac{8\pi}{3\beta}}\left(  \varphi+C\right)  \right)  \right)  ,
\label{18}%
\end{equation}
Taking properly the constant of integration $C$ and expanding this potential
for large $\varphi$%
\begin{align}
V\left(  \varphi\right)   &  \simeq V_{0}\left[  \exp\left(  -\frac{3\beta}%
{2}\exp\left(  -\sqrt{\frac{8\pi}{3\beta}}\left(  \varphi+C\right)  \right)
\right)  \right]  ^{2}\nonumber\\
&  \simeq V_{0}\left[  1-\exp\left(  -\sqrt{\frac{8\pi}{3\beta}}%
\varphi\right)  \right]  ^{2}, \label{19}%
\end{align}
we immediately recognize, for $\beta=1/2,$ the $R^{2}-$ inflation
\cite{StarMukh} and Higgs inflation \cite{BS}, which are indistinguishable
experimentally. They predict not only the same spectral index but also exactly
the same amount of the gravitational waves $r=12/\left(  N+1\right)
^{2}\simeq5\times10^{-3}.$ The other models in this class correspond to
different values of $\beta$ and can be distinguished only by the amount of the
produced gravitational waves.

$\alpha\neq1,2$

In this case the potential is%
\begin{equation}
V\left(  \varphi\right)  \simeq V_{0}\exp\left[  -\frac{3\beta}{\alpha
-1}\left(  \frac{2\pi}{3\beta}\left(  2-\alpha\right)  \varphi^{2}\right)
^{\frac{\alpha-1}{\alpha-2}}\right]  . \label{20}%
\end{equation}
First of all note that if $\alpha=0$ then $1+p/\varepsilon=\beta$ and for
$\beta\ll1$ we have inflation which continues forever without graceful exit.
As one can see from (\ref{20}) this case corresponds to the exponential
potential $V\left(  \varphi\right)  \simeq V_{0}\exp\left(  \sqrt{12\pi\beta
}\varphi\right)  .$ To understand the difference between the cases
$0<\alpha\,<1,$ $1<\alpha\,<2$ and $\alpha>2$ we will consider separately
$\alpha=1/2,3/2$ and $\alpha=3.$ As one can see from (\ref{12}) the first two
cases correspond, as before, to large field inflation, while in the third case
we have small field inflation.

For $\alpha=1/2$ the potential becomes%
\begin{equation}
V\left(  \varphi\right)  \simeq V_{0}\exp\left(  6\pi^{1/3}\left(
\beta\varphi\right)  ^{2/3}\right)  \label{21}%
\end{equation}
and for $\varphi>1$ it satisfies the slow roll conditions. For instance, in
case when $3\beta=1,$ the spectral index $n_{s}\simeq0.85$ and $r\simeq1.13$
are in obvious contradiction with observations.

The case $\alpha=3/2$ corresponds to the potential%
\begin{equation}
V\left(  \varphi\right)  \simeq V_{0}\exp\left(  -\frac{27\beta}{\pi
\varphi^{2}}\right)  \simeq V_{0}\left(  1-\frac{27\beta}{\pi\varphi^{2}%
}\right)  , \label{22}%
\end{equation}
for $\varphi>1,$ where this potential is trustable. The situation here is
similar to $\alpha=2$ case$,$ where \ for large $\varphi$ the potential
approaches a constant value exponentially fast. Here this constant is
approached only as an inverse power law. The spectral index $n_{s}\simeq0.967$
and $r\simeq2\times10^{-2}$ in good agreement with observations$.$ It seems
that taking into account the possible accuracy of the measurements, the
uncertainty in $N$ due to unknown detailed physics after inflation and the
remaining freedom in the choice of $\beta$ one will never be able to
distinguish the models described by potentials of type (\ref{18}) and
(\ref{22}) even if one will ever measure the gravitational waves. On the other
hand, the models with potentials as in (\ref{13}) and (\ref{21}) seems could
soon be ruled out even if one only improve the accuracy of the determination
of the spectral index and the upper bound on $r$, for instance, in Planck
experiment. Therefore, although non-decisive for selecting a particular
scenario the precision measurements are very useful for excluding the whole
families of inflationary scenarios.

Finally let us consider the case of $\alpha=3,$%
\begin{equation}
V\left(  \varphi\right)  \simeq V_{0}\exp\left(  -\frac{2\pi^{2}}{3\beta
}\varphi^{4}\right)  \simeq V_{0}\left(  1-\frac{2\pi^{2}}{3\beta}\varphi
^{4}\right)  \label{33}%
\end{equation}
Here inflation occurs at small values of scalar field$.$ The spectral index
corresponds to $n_{s}\simeq0.94$ and the amount of the gravity waves is
further suppressed by $N$ compared to $\alpha=2$ case, namely, $r=24\beta
/\left(  N+1\right)  ^{3}\simeq6\times10^{-5}$ for $3\beta=1.$

The potentials above approximate the inflationary potential only in the range
of slow roll regime, or even more conservatively within last 70 e-folds of
inflation. It is clear that outside this range they can be changed rather arbitrarily.

\section{k-Inflation}

Until now we have characterized inflation only by one function, namely, by
$N$-dependence of deviation of the equation of state from cosmological
constant. However, yet remaining in one fluid approximation one can introduce
the other natural parameter which describes this fluid, namely, the speed of
propagation of cosmological perturbations, or in other words, the speed of
sound $c_{s},$ which naturally arises in k-inflation \cite{ADM}. At present to
explain the observations there is no need to assume that $c_{s}$ is different
from unity. However, because slow roll and k-inflation scenarios belong to the
same class of simplicity it is interesting to check how this extra parameter
will influence the robust predictions above. In case of $c_{s}\neq1$ the
formulae (\ref{3}) and (\ref{7}) are modified as \cite{Mbook}:%
\begin{equation}
n_{s}-1=-3\left(  1+\frac{p}{\varepsilon}\right)  +\frac{d}{dN}\ln\left[
c_{s}\left(  1+\frac{p}{\varepsilon}\right)  \right]  , \label{34}%
\end{equation}
and
\begin{equation}
r=\frac{T}{S}=24c_{s}\left(  1+\frac{p}{\varepsilon}\right)  . \label{35}%
\end{equation}
We can parametrize k-inflation with%
\begin{equation}
1+\frac{p}{\varepsilon}=\frac{\beta}{\left(  N+1\right)  ^{\alpha}},\text{
\ }c_{s}=\frac{\gamma}{\left(  N+1\right)  ^{\delta}}, \label{36}%
\end{equation}
where $\delta\geq0$ because the speed of sound generically grows towards the
end of inflation\cite{ADM}, and $\gamma$ is now an arbitrary positive number
not necessarily of order unity. The expression for the spectral index becomes%
\begin{equation}
n_{s}-1=-\frac{3\beta}{\left(  N+1\right)  ^{\alpha}}-\frac{\alpha+\delta
}{\left(  N+1\right)  }, \label{37}%
\end{equation}
and because $\delta$ is positive the conclusion about minimal deviations of
the spectral index from unity remains unaltered. On the other hand it looks
like the tensor to scalar ratio%
\begin{equation}
r=\frac{24\beta\gamma}{\left(  N+1\right)  ^{\alpha+\delta}}, \label{38}%
\end{equation}
can be made arbitrarily small because $\gamma$ is not needed to be of order
unity and as a result the lower bound on the amount of gravitational waves
will disappear. However, it is not so. In fact, too small speed of sound
induces too large primordial non-gaussianities of order $f_{NL}\simeq O\left(
1\right)  /c_{s}^{2}$. Therefore taking the experimental bound on $f_{NL}<80$
\ \ \cite{komatsu}, we conclude that the speed of sound cannot be much smaller
than $0.1$ and hence for the spectral index $n_{s}\simeq0.96$ the lower bound
on $r$ (see (\ref{7c})) is not modified more than by an order of magnitude at maximum.

\section{Discussion}

We have shown that the majority of inflationary scenarios can be parametrized
by two numbers and using this we were able to prove the bound on the minimal
deviations of spectral index from unity and to obtain the lower bound on the
amount of the gravitational waves produced. Although primordial gravitational
waves are not yet detected, the experimental confirmation of the flatness of
the universe, adiabatic nature of nearly gaussian perturbations and the
discovered (at 3,5 sigma level) logarithmic tilt of the spectrum unambiguously
prove the quantum origin of the universe structure and the early cosmic
acceleration. Needless to say that all these predictions, which were yet in
conflict with observations about 15 years ago, are very nontrivial. Given that
the quantum origin of the universe structure is experimentally confirmed, the
precision measurements already now allow us to exclude many inflationary
scenarios existing in the literature. Moreover, the improved accuracy of the
determination of spectral index, the bound (or detection) on non-gaussianity
and the bound (or possible future detection) on primordial gravitational waves
will allow us to put further restrictions on the admissible inflationary
scenarios. However, this seems will not help us too much in recovering the
fundamental particle physics behind inflation. In fact, the observational data
only allow us to measure only the effective equation of state and the rate of
its change in a rather small interval of scales. Keeping in mind unavoidable
experimental uncertainty, the effect of unknown physics right after inflation
and degeneracy in the scenarios discussed above we perhaps will never be able
to find out the microscopical theory of inflation without further very
essential input from the particle physics. On the other hand, the remarkable
property of the theory of quantum origin of the universe structure is that the
gravity seems does not care too much about microscopic theory providing needed
equation of state, and allows us to make experimentally verifiable predictions.

\textbf{Acknowledgements} I am grateful to Cesar Gomez for useful discussions.
This work was supported by \textquotedblleft Chaire Internationale de
Recherche Blaise Pascal financ\'{e}e par l'Etat et la R\'{e}gion
d'Ile-de-France, g\'{e}r\'{e}e par la Fondation de l'Ecole Normale
Sup\'{e}rieure\textquotedblright, by TRR 33 \textquotedblleft The Dark
Universe\textquotedblright\ and the Cluster of Excellence EXC 153
\textquotedblleft Origin and Structure of the Universe\textquotedblright.

\end{document}